\documentclass[prd,twocolumn,floatfix,preprintnumbers,reprint,showpacs]{revtex4}
\usepackage{epsfig,amsmath,amssymb,bm,color,graphicx}
\usepackage{afterpage}
\usepackage{pdfpages}
\def\lya{Lyman-$\alpha$\,}
\def\mWDM{m_{\rm WDM}}

\newcommand{\skm}{\mathrm{s\,km^{-1}}}
\newcommand{\kms}{\mathrm{km\,s^{-1}}}

\newcommand{\be}{\begin{equation}}
\newcommand{\ee}{\end{equation}}

\usepackage{color}

\begin{document}

\title{New Constraints on the free-streaming of warm dark matter from intermediate and small scale \lya forest data}
\author{Vid Ir\v{s}i\v{c}$^{1,2,3}$}\thanks{E-mail: irsic@uw.edu (VI)}
\author{Matteo Viel$^{4,5,6}$} \thanks{E-mail: viel@sissa.it (MV)}
\author{Martin G. Haehnelt $^{7}$}
\author{James S. Bolton $^{8}$}
\author{Stefano Cristiani$^{5,6}$}
\author{George D. Becker$^{7,9}$}

\author{Valentina D'Odorico$^{5}$}
\author{Guido Cupani$^{5}$}
\author{Tae-Sun Kim$^{5}$}
\author{Trystyn A. M. Berg$^{10}$}
\author{Sebastian L{\'o}pez$^{11}$}
\author{Sara Ellison$^{10}$}

\author{Lise Christensen$^{12}$}
\author{Kelly D. Denny$^{13}$}
\author{G\'{a}bor Worseck$^{14}$}

% List of institutions
\smallskip
\affiliation{
$^{1}$University of Washington, Department of Astronomy, 3910 15th Ave
NE, WA 98195-1580 Seattle, USA\\
$^{2}$Institute for Advanced Study, 1 Einstein Drive, NJ 08540 Princeton, USA\\
$^{3}$The Abdus Salam International Centre for Theoretical Physics,
Strada Costiera 11, I-34151 Trieste, Italy\\
$^{4}$ SISSA-International School for Advanced Studies, Via Bonomea 265, 34136 Trieste, Italy \\
$^{5}$INAF - Osservatorio Astronomico di Trieste, Via G. B. Tiepolo
11, I-34143 Trieste, Italy\\
$^{6}$INFN - National Institute for Nuclear Physics, via Valerio 2,
I-34127 Trieste, Italy\\
$^{7}$ Institute of Astronomy and Kavli Institute of Cosmology, Madingley Road, Cambridge CB3 0HA, UK \\
$^{8}$ School of Physics and Astronomy, University of Nottingham, University Park, Nottingham, NG7 2RD, UK\\
$^{9}$Space Telescope Science Institute, 3700 San Martin Drive,
Baltimore, MD 21218, USA\\
$^{10}$Department of Physics and Astronomy, University of Victoria,
Victoria, BC V8P 1A1, Canada\\
$^{11}$Departamento de Astronom\'{i}a, Universidad de Chile, Casilla
36-D, Santiago, Chile\\
$^{12}$Dark Cosmology Centre, Niels Bohr Institute, University of
Copenhagen, Juliane Maries Vej 30, DK-2100 Copenhagen, Denmark\\
$^{13}$Department of Astronomy, The Ohio State University, 140 West
18th Avenue, Columbus, OH 43210, USA\\
$^{14}$Max-Planck-Institut f\"{u}r Astronomie, K\"{o}nigstuhl 17,
D-69117 Heidelberg, Germany \\
}

\begin{abstract}
We present new measurements of the free-streaming of warm dark matter
(WDM) from \lya flux-power spectra.  We use data from the medium
resolution, intermediate redshift XQ-100 sample observed with the
X-shooter spectrograph ($z=3 - 4.2$) and the high-resolution, high-redshift sample
used in Viel et al. (2013) obtained with the HIRES/MIKE
spectrographs ($z=4.2 - 5.4$). Based on further improved modelling of the dependence
of the \lya flux-power spectrum on the free-streaming of dark matter,
cosmological parameters, as well as the thermal history of the
intergalactic medium (IGM) with hydrodynamical simulations, we obtain
the following limits, expressed as the equivalent mass of thermal
relic WDM particles. The XQ-100 flux power spectrum alone gives a
lower limit of 1.4 keV, the re-analysis of the HIRES/MIKE sample gives
4.1 keV while the combined analysis gives our best and significantly
strengthened lower limit of 5.3 keV (all 2$\sigma$ C.L.). The further
improvement in the joint analysis is partly due to the fact that the
two data sets have different degeneracies between astrophysical and
cosmological parameters that are broken when the data sets are
combined, and more importantly on chosen priors on the thermal
evolution. These results all assume that the temperature evolution of
the IGM can be modelled as a power law in redshift. Allowing for a
non-smooth evolution of the temperature of the IGM with sudden
temperature changes of up to 5000K reduces the lower limit for the
combined analysis to 3.5 keV. A WDM with smaller thermal relic masses
would require, however, a sudden temperature jump of 5000\,K or more
in the narrow redshift interval $z=4.6-4.8$, in disagreement with
observations of the thermal history based on high-resolution
resolution \lya forest data and expectations for photo-heating and
cooling in the low density IGM at these redshifts.
\end{abstract}

\maketitle
\section{Introduction}

The intergalactic medium (IGM) and its main observable manifestation,
the \lya\ forest (see \citep{mcquinn15}), have been used as unique
tools to address key cosmological issues: the free-streaming of
dark matter and in particular departures from cold dark matter,
generally labelled as warm dark matter (WDM) 
\cite{nara00,viel05,uros06,viel08,bird10,viel13WDM,baur15}; measuring the
linear power spectrum at small scales
\citep{croft02,zaldarriaga03,mcdonald03,viel04,mcdonald05,seljak06};
probing the effect of the free-streaming and thus the masses of
neutrinos \citep{seljak06,palanque15}, and placing (high-redshift)
geometrical constraints on our Universe from Baryonic Acoustic
Oscillations measurements \citep{busca13,slosar13}.

At present, constraints on the matter power spectrum are either
derived from moderate size samples with tens of high-resolution, high
signal-to-noise spectra (VLT, HIRES/KECK,
\citep{croft02,viel04,viel13WDM}) or large samples with thousands of
low-resolution, low signal-to-noise spectra (SDSS-II, SDSS-III/BOSS,
\citep{mcdonald05,boss13,palanque13}). The XQ-100 \citep{lopez16}
sample bridges the gap between these two regimes with its homogeneous
set of intermediate resolution and intermediate signal-to-noise QSO
absorption spectra, with the additional benefit that the flux power
spectrum inferred from medium resolution QSO absorption spectra is
subject to quite different systematic and statistical uncertainties.
Here, we will provide constraints on the free-streaming length of dark
matter from modelling the XQ-100 flux-power spectrum as well as from a
combined analysis with new modelling of the HIRES/MIKE flux power
spectrum presented in \citep{viel13WDM}. These two data sets have a
small redshift overlap and it can be expected that a combined analysis
will further break degeneracies with remaining uncertainties in the
parameters describing the thermal evolution of the IGM, the evolution
of the mean flux and cosmological parameters. Pushing the constraints
on the free-streaming length of dark matter as far as possible is very
relevant for the wider astrophysics community given that considerable
tensions with the Cold Dark Matter (CDM) model on small scales
continue to persist for a range of astrophysical observations,
especially with regard to the dynamical properties of Milky Way
satellites (see e.g \citep{weinberg15}). From a particle physics point
of view, small scale modifications of the cold dark matter power
spectrum can {\it e.g.} arise from the free-streaming of sterile
neutrinos \citep{whitepaper,bulbul16,boyarsky14} or ultra-light bosons \citep{hui16},
alternatives to the more generic thermal relics on which we
concentrate our modelling here \citep{bode01}. For the purpose of
our analysis here the different DM candidates differ in the exact
shape of the suppression of the small scale power spectrum they cause
due to free-streaming. However, apart from changing the nature of the
DM several studies have shown that baryonic physics could help in
alleviating or even solving the small scale tensions \citep{brooks14,wetzel16}.

In Section~\ref{datasec} we briefly describe the data sets used;
Section~\ref{simsec} presents the hydrodynamical simulations grid; the
method is briefly outlined in Section~\ref{methodsec} while
Section~\ref{ressec} contains all the new results (with an Appendix
focusing on degeneracies between the various parameters).  We conclude
with a summary in Section ~\ref{conclusec}.

\section{Data sets} 
\label{datasec}

We make use of two different and complementary data sets: the XQ-100
and HIRES/MIKE samples. XQ-100 consists of 100 medium resolution and
signal-to-noise QSO spectra obtained as part of the XQ-100 survey with
emission redshifts $3.5 < z < 4.5$ \citep{lopez16}. A detailed
description of the data and the 1D flux power spectrum measurements is
presented in \citep{irsic17}. The resolution of the XQ-100 spectra is
$30-50 \kms$ (FWHM) depending on wavelength and the pixel size used for
calculating the flux power spectrum for the two spectral arms is $20$ (UVB)
and $11\;\kms$ (VIS), respectively (see \citep{irsic17} for details).

The spectral resolution sets the smallest scales probed by the data.
The flux power spectrum extraction has been extensively modeled with
mock data sets built from hydrodynamic simulations which allows an
accurate estimation of statistical and systematic uncertainties of the
flux power at $z=3,3.2,3.4,3.6,3.8,4,4.2$ for 19 bins in $k-$space in
the range $0.003$--$0.057\rm\, \skm$. In \citep{irsic17} the
covariance matrix was multiplied with a constant factor 1.1, to
correct for the underestimation of variance through the bootstrap
method.  We use the same correction factor here unless otherwise
noted. We further assume that the cross-correlations between different
redshift bins is zero. A total of 133 $(k,z)$ points are thus used in
our modelling of the XQ-100 flux power spectrum.

We also combine the new data with measurements of the power spectrum
used in \citep{viel13WDM}, measured at higher redshift
$z=4.2,4.6,5.0,5.4$, for 10 $k-$bins in the range $0.001$--$0.08\rm\,
\skm$. The QSO absorption spectra of this sample have been
obtained with the HIRES/KECK and the MIKE/Magellan spectrographs, with
resolution 6.7 $\kms$ and 13.6 $\kms$, respectively. Following
\cite{viel13WDM} a conservative cut is imposed on the MIKE and HIRES
data, such that wavenumbers with $k < 0.005\;\skm$ are removed, due to possible systematic
uncertainties on the large scales of those measurements that used only
a small number of QSO sightlines. Moreover, we also do not consider
the highest redshift bin for the MIKE data, where the flux power
spectrum measurements have large error bars. The analysis in
\cite{viel13WDM} used a correction factor of $1.5$ for the nominal
error estimates of the flux power spectrum. We use the same value in
our analysis here, unless otherwise stated. Ref.~ \cite{viel13WDM}
further checked the error estimates using a sample of mock absorption
spectra. This suggested that the two error bars in the MIKE and HIRES
data sets with $\sigma_F/P_F < 0.075$ should be set to 0.075, which we
also do in our analysis here (see \citep{viel13WDM} for more details).
We further regularize the covariance matrix of the HIRES/MIKE data
following the procedure of \cite{viel13WDM}. A total of 49 $(k,z)$
points is used in the HIRES/MIKE analysis.

\section{Simulations} 
\label{simsec}

Our analysis of the flux power spectrum is based on a set of
hydrodynamical simulations that is significantly extended compared to
that used in \cite{viel13WDM}. The hydrodynamical simulations were
performed with the GADGET-3 code, which is a modified version of the
publicly available GADGET-2 code \citep{gadget}. A simplified star
formation criterion is applied for which gas particles above an
overdensity 1000 and temperature below T$=10^5$ K are converted into
stars (e.g. \citep{bolton16}). The reference model simulation has a
box length of 20$/h$ comoving Mpc with $2 \times 768^3$ gas and (cold)
dark matter particles (with gravitational softening 1.04$/h$ comoving
kpc) in a flat $\Lambda$CDM universe with cosmological parameters
$\Omega_{\rm m} = 0.301$, $\Omega_{\rm b} = 0.0457$, $n_{\rm s} =
0.961$, $H_0 = 70.2\;\mathrm{km\,s^{-1}\,Mpc^{-1}}$ and $\sigma_8 =
0.829$ in agreement with \citep{planck15}. Three different WDM models
with masses $m_{\rm WDM} = 2,3,4\;\mathrm{keV}$ have also been
simulated. Initial conditions were implemented using the same approach
as \citep{viel05}.  We explore the thermal history of the \lya\ forest
by modifying the photo-heating rates in the simulations as in
\citep{bolton08}. The low density IGM ($\Delta=1+\delta < 10$) is well
described by a power-law temperature-density relation, $T =
T_0\Delta^{\gamma-1}$. We consider a range of values for the
temperature at mean density $T_0$ and the slope of the $T-\rho$
relation, $\gamma$, based on the previous analysis of the \lya\ forest
and recent observations \citep{becker11}. These consist of a set of
three different temperatures at mean density, $T_0(z=3.6) = 7200,
11000, 14800\;\mathrm{K}$, which evolve with redshift, as well as a
set of three values of the slope of the $T-\rho$ relation:
$\gamma(z=3.6) = 1.0, 1.3, 1.5$. These 9 thermal history models have
been simulated for the reference $\Lambda$CDM case; additionally all
three different temperature models have been simulated for all three
WDM models as well. The reference thermal history assumes
$(T_0(z=3.6),\gamma(z=3.6)) = (11000\;\mathrm{K},1.5)$.

Instead of using the standard cosmological parameters of $\sigma_8$,
the slope of the initial power spectrum $n_{\rm s}$ and $\Omega_{\rm
  m}$ as in \cite{viel13WDM}, we exploit the fact that these three
parameters are tightly connected in $\Lambda$CDM (and $\Lambda$WDM)
models and impact on the flux power spectrum only in terms of the
amplitude and the (effective) slope of the {\it matter} power spectrum
{\it at scales that are probed by the \lya\ forest}. We therefore use
instead only two parameters describing cosmology, $\sigma_8$ and
$n_{\rm eff} = d\ln{P_{\rm m}(k)}/d\ln{k}$, evaluated at $k =
0.005\;\skm$, similarly to what was done in \cite{mcdonald05}. Five
different values are considered for both $\sigma_8 =
0.754,0.804,0.829,0.854,0.904$, and $n_{\rm eff} =
-2.3474,-2.3274,-2.3074,-2.2874,-2.2674$. The reference model has
$(\sigma_8,n_{\rm eff}, n_{\rm s}) = (0.829,-2.3074,0.961)$.  Other
cosmological parameters are kept fixed at the Planck best fit
values. In practice, the change in $n_{\rm eff}$ is implemented with
small changes of $n_{\rm s}$.

We also vary the redshift of reionization $z_{\rm rei}$ which is
chosen to be $z_{\rm rei} = 9$ for the reference model as well as
$z_{\rm rei} = 7,15$ for two additional models. The $z_{\rm rei}=7$
model has also been simulated for all 3 values of the WDM thermal
relic mass, since the redshift of reionization has an impact on the
Jeans smoothing scale and could affect the cutoff scale of the flux
power spectrum. We note here, however, that the effect is large enough
for the degeneracy between free-streaming and Jeans smoothing to be
broken (although see \citep{nasir16} in
the context of CDM models) and that the data are not constraining this
parameter well (see \ref{app:B} for details).

The final parameter we explored characterizes the possible effect of
ultraviolet (UV) background fluctuations. A model has been chosen
where the spatial fluctuations of the meta-galactic UV background are
dominated by rare QSOs, which has a strong scale dependent effect on
the flux power spectrum particularly at high redshift and at large
scales. The model of UV fluctuations used here is an update of the
model presented in \citep{viel13WDM} (see appendix there). The updated
model uses the more recent mean free path measurements of
\citep{worseck14} and parameterizes the effect of UV fluctuations on
the flux power spectrum as $f_{\rm UV}$ -- defined as the fraction of
the volume averaged hydrogen photo-ionisation rate that arises from a
fluctuating QSO component. The remaining fraction, $1-f_{\rm UV}$ is
attributed to a spatially uniform UV background arising from faint
galaxies with a typical separation much less than the mean free path
of ionising photons.  The flux power spectrum template is built from a
set of 3 models variations with $f_{\rm UV}=0,\,0.5,\,1$ where $f_{\rm
  UV}=0$ corresponds to a spatially uniform UV background.  Note,
however, that a comprehensive treatment of spatial UV (and
temperature, which we neglect here) fluctuations would require
computationally prohibitive radiative transfer calculations in large
volumes. As discussed in \cite{hui16} spatial variations in the IGM
temperature \citep{daloisio15}, mean free path \citep{becker15} and
fluctuations from bright Lyman break galaxies at high redshift
\citep{mcdonaldwinds} (particularly at $z>5$) may also have an
uncertain impact on the flux power.

Last but not least, we also vary the mean flux (or equivalently amplitude of the
UV background) by rescaling $\tau_{\rm eff}=-\ln{\bar F}$.  We use three different
values $(0.8,1,1.2) \times \tau_{\rm obs, eff}$, with the reference
values of $\tau_{\rm obs,eff}$ chosen to be those of the SDSS-III/BOSS
measurements \citep{palanque13}. The mean flux evolution derived from
the SDSS-III/BOSS analysis has values that are 5-8\% lower compared to
those measured by \citep{becker11}, but note that the range of values
considered in our analysis brackets the observed values by
\citep{becker11} as well.

Finally, a few lower resolution simulations have also been run to
check convergence and a single 1 keV WDM model has been considered to
check the validity of the method described below.  Each simulation
used about 20,000 CPU hours. The total grid consists of 23
simulations at the reference resolution and 10 simulations at lower
resolution.

\section{Method}
\label{methodsec}

Using the models of the transmitted flux obtained from the simulations
we establish a grid of points for each redshift, in the parameter
space of $({\bar F}(z), T_0(z), \gamma(z), \sigma_8, z_{\rm rei},
n_{\rm eff}, f_{\rm UV}, m_{\rm WDM})$. We then perform a linear
interpolation between the grid points in this multidimensional
parameter space. The interpolation is done in the $P_{\rm F}(k,z)$
space directly, rather than for ratios of flux power spectra as in
\citep{viel13WDM}. We perform several tests of the interpolation
scheme (by predicting the value of the flux power at a given grid
point where exact values are known, without using that grid point in
the interpolation) and conclude that while a small systematic error
due to interpolation exists ($<5\%$ of the flux power spectrum), 
it does not bias the results. Additional tests were done when
including this correction in the error budget of the likelihood
estimation and results were unchanged. This reflects the fact that
the interpolation error is small compared to the statistical error and
sub-dominant in the systematic error budget of the current data. 
A Gaussian likelihood estimation was
then used to evaluate a Monte Carlo Markov Chain (MCMC) algorithm to
obtain the set of parameters that minimizes the likelihood for a given
data set. 

To estimate the convergence, four independent chains were run from
randomly chosen initial set of parameters with different seed values
for pseudo-random number generators. Using the Gelman-Rubin test on
all of the chains we concluded that the chains have converged
sufficiently (for each of the parameters the Gelman-Rubin measure of
convergence was required to be less than $1.1$). The resulting chains
were combined after pruning the burn-in samples, from which the
estimates of the posterior distributions (and its moments) were
obtained.

\begin{figure*}
\begin{center}
\includegraphics[width=15cm]{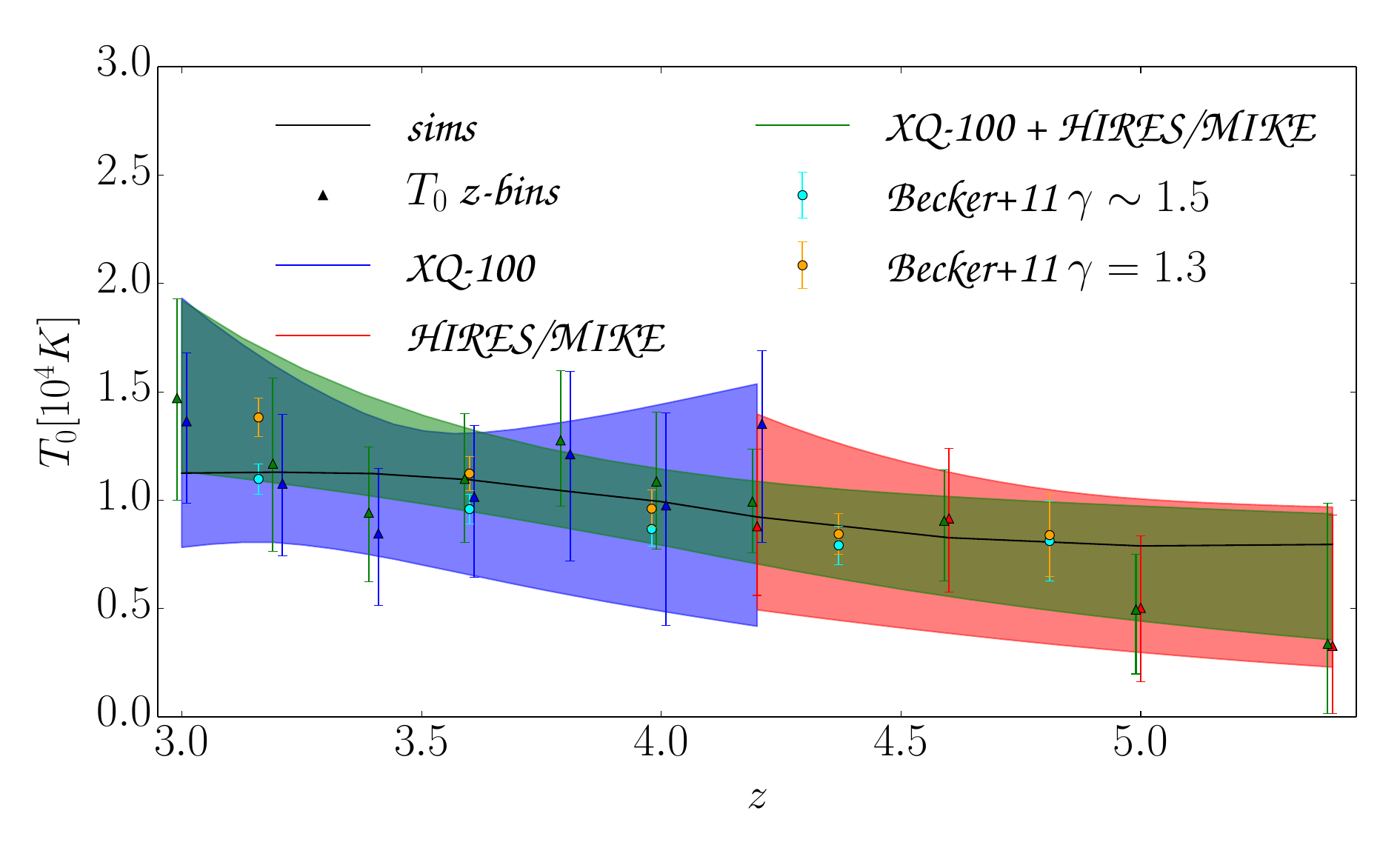}
\caption{Temperature measurements ($2\sigma$) as a function of redshift:
  reference simulation (black curve), XQ-100 (shaded blue area),
  HIRES/MIKE (shaded red area), joint constraints (shaded green area);
  points (same color coding) represent measurements obtained using
  T$_0$ in redshift bins, limiting temperature variations to  $\Delta
  T=5000\rm\,K$ between adjacent bins, rather than assuming a
  power-law evolution. Cyan and orange points with error bars are the
  IGM temperature measurements from \citep{becker11} for two values
  of the slope of the temperature-density relation, $\gamma=1.3$ and
  $1.5$.}
\label{fig_2} 
\end{center}
\end{figure*}

\section{Results}
\label{ressec}

We performed a detailed MCMC analysis for three different data sets:
XQ-100 (the new data set), HIRES/MIKE (as in \citep{viel13WDM}) and
the combined data sets. For the reference analysis case we model the
mean flux parameters independently for each redshift bin, the number
of which varies for each data set (XQ-100 has $7$, MIKE/HIRES has $4$
and the combined analysis has $10$ redshift bins). We complement these
parameters with an additional $9$ parameters: $5$ parameters
describing either cosmology or astrophysics ($\sigma_8$, $n_{\rm
  eff}$, $z_{\rm rei}$, $m_{\rm WDM}$, $f_{\rm UV}$) and $4$
parameters describing the thermal state of the IGM, using a power-law
$T-\rho$ relation, $T = T_0\Delta^{\gamma-1}$. Unless otherwise noted
we model the redshift evolution of the parameters $T_0$ and $\gamma$
as power-laws, such that $T_0(z) = T_0^A\left[(1+z)/(1+z_{\rm
    p})\right]^{T_0^S}$ and $\gamma(z) = \gamma^A\left[(1+z)/(1+z_{\rm
    p})\right]^{\gamma^S}$. The pivot redshift is different for each
data set and roughly corresponds to the redshift at which most of the
\lya\ forest pixels are coming from ($z_{\rm p} = 3.6, 4.5, 4.2$ for
XQ-100, HIRES/MIKE and the combined analysis, respectively).  As we
will see later the choice of thermal priors significantly affects the
results.

A default set of priors was used in the reference runs for the mean
flux (${\bar F}$) in each redshift bin, according to the $\tau_{\rm
  eff}$ fit to the data presented in \citep{viel13WDM}, with $0.04$
standard deviation (1$\sigma$). These priors account for the fact that
different continuum treatments and different measurements give a
slightly different normalization for the mean flux. The chosen fit
roughly represents the median values of the observations (see
\citep{irsic17}), with the 1$\sigma$ standard deviations capturing the
uncertainty in the normalization given by different measurements.

For all three data sets, the preferred ranges of other parameters are
in agreement with independent observations. In particular the values
of cosmological parameters $\sigma_8$ and $n_{\rm eff}$ are consistent
with the latest Planck results within 1$\sigma$ for XQ-100 and
HIRES/MIKE and within 2$\sigma$ for the combined analyses of XQ-100
and HIRES/MIKE. We have furthermore verified that the moderate
2$\sigma$ discrepancy in $\sigma_8$ (and to lesser extent in $n_{\rm
  eff}$) can be alleviated by using additional priors on the above
parameters. The applied priors were Gaussian on $\sigma_8$ and n$_{\rm eff}$ of $\pm$
0.01 (1$\sigma$) around Planck values. Our measurements of the
cosmological parameters are consistent with those measured by
SDSS/BOSS collaboration \citep{palanque13}, and moreover also show
a similar tendency towards slightly higher values of $\sigma_8$ and
slightly lower values of $n_{\rm eff}$.

It is also important to emphasise that the redshift coverage of XQ-100
and the higher resolution HIRES/MIKE data sets is mostly complementary
(covering lower and higher redshifts respectively) and thus different
constraints and degeneracies are expected in each.  Even though XQ-100
covers a similar redshift range as the SDSS-II and SDSS-III
\lya\ power spectrum measurements, it extends to significantly smaller
scales and should carry more information from the thermal cut-off in
the flux power-spectrum.  Note that the thermal cut-off is fixed in
comoving co-ordinates in real space, while the cut-off in the observed
transmitted flux power spectrum scales as $H(z)/(1+z)$ in velocity
space. At a fixed velocity scale this means smaller comoving length
scales (and thus free streaming lengths) are probed with increasing
redshift. As a result, higher redshift data are more sensitive to the
equivalent larger WDM relic mass than lower redshift data, where the
effect of the thermal motions dominates already at larger comoving
length scales.  Measurements of WDM from lower redshift data, like
those obtained from SDSS/BOSS flux power spectra, are thus mostly
sensitive to the change of the power spectrum amplitude on the large
scales, instead of probing the shape and redshift evolution of the
free-streaming cut-off.  In our analysis, this is supported by the
fact that large degeneracies are found in our MCMC analysis for XQ-100
between the WDM mass and the values of the mean flux at each redshift
(see Fig.~\ref{fig_WDM} in the appendix).  Furthermore, since XQ-100
consists of fewer QSO spectra, the error bars are larger than that of
the SDSS measurements, which is why we do not expect the results from
XQ-100 alone to constrain the WDM mass as tightly as various SDSS
measurements.

Before discussing our new free-streaming constraints, in
Fig.~\ref{fig_2} we show the temperature estimates from our MCMC
analysis of the flux power spectrum for the different data sets as
2$\sigma$ shaded regions, assuming the temperature of the IGM varies
smoothly with redshift as a power-law.  In addition, we show
individual points with 2$\sigma$ errorbars that are obtained by
allowing the temperature to float freely from bin to bin, but with a
maximum temperature jump between bins of $\Delta T= 5000\rm\,K$
(discussed further below).  Both results are in good agreement with
the measurements of \cite{becker11} obtained from the curvature of the
transmitted flux, shown as orange and cyan points for two different
assumptions for the power-law slope $\gamma$ of the
temperature-density relation (but note that these measurements were
calibrated with hydrodynamical simulations where the dark matter was
assumed to be cold).  While our measurements are consistent with no
evolution in temperature in the redshift range $3 < z < 5.4$, the
preferred slope is negative (temperature increasing with decreasing
redshift), which is in agreement with HeII reionisation occuring
somewhere around redshifts $3 - 4$ as inferred from high-resolution
\lya forest data \citep{schaye00,lidz10,puchwein15,upton}.

Earlier measurements of the IGM temperature at $z=3$--$4.3$ by
\citep{schaye00} using a Voigt profile fitting approach are also in
reasonable agreement with our XQ-100 constraint (blue shading),
although the error bars are large in that study.  More recently,
Ref.~\citep{lidz10} measured $T_{0}\simeq 26,\,000\pm 5000\rm\,K$
($2\sigma$) at $z=3.4$ using wavelets, a result which is in greater
tension with our measurement.  This suggests there are still some
systematic differences between these analyses likely associated with
the calibration of hydrodynamical simulations and observational data.
Note also that higher gas temperatures will in general raise the lower
limit on the WDM particle mass and tighten the constraint further, so
in this sense our constraint can be considered conservative.

\begin{figure*}
%\begin{center}
\includegraphics[width=15cm]{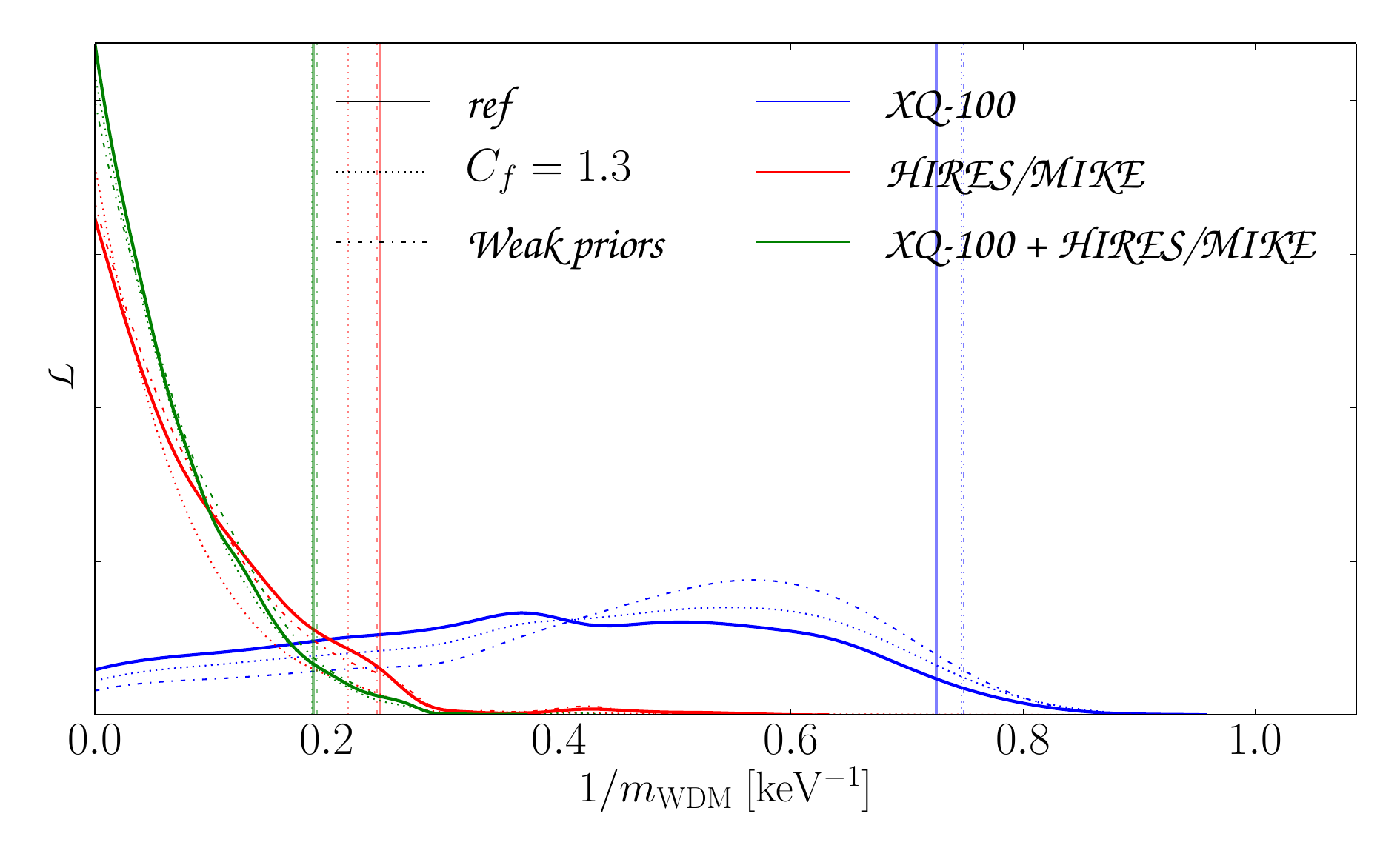}
\caption{One dimensional posterior likelihood distributions for the
  WDM mass for XQ-100, HIRES/MIKE and the combined data (blue, red and
  green solid curves).  We also show how the results change by using a
  larger value for the correction factor of the XQ-100 covariance
  matrix (dotted curves) and using weak priors (see text) on the
  thermal history and cosmological parameters (dot-dashed
  curves). Vertical lines show the corresponding 2$\sigma$ confidence
  limits.}
\label{fig_3}
%\end{center} 
\end{figure*}

Table 1 presents our marginalized parameter constraints, and
Fig.~\ref{fig_3} shows our constraints on the free-streaming of dark
matter expressed as the mass of a putative thermal relic WDM particle.
As expected the results from the XQ-100 data set alone (solid blue
curve) only weakly constrain the mass ($m_{\rm WDM} >
1.4\;\mathrm{keV}$ at 2$\sigma$).  While the peak of the likelihood is
not at $0$, the peak is not statistically significant (not even at
1$\sigma$).  Moreover, the exact position of the peak is strongly
dependent on the choice of priors.  However, the 2$\sigma$ upper limit
for $1/m_{\rm WDM}$ is nearly independent of prior choice, and
constitutes a very robust measurement.  We also show the case where a
correction factor of 1.3 has been applied to the covariance matrix and
with weak priors on cosmological parameters ($\sigma_8$ and n$_{\rm
  eff}$ have Gaussian priors of $\pm$ 0.1 (1$\sigma$) around Planck
values and the assumed temperature $T_0^A$ is $10,000 \pm 5000$ K
(1$\sigma$)).  When we move to the model with with freely floating
$T(z)$ bins rather than a power-law evolution of the temperature the
free-streaming length inferred from the XQ-100 sample does not change.

Constraints on the WDM mass using the HIRES/MIKE sample were first
presented in \citep{viel13WDM}.  Compared to the analysis presented in
\cite{viel13WDM} the main improvements in this work are as follows:
the reference simulations have higher resolution and better coverage
of parameter space, the model of spatial UV fluctuations has been
extended, the interpolation scheme is based on the prediction of flux
power rather than flux power ratios, and the (now one) cosmological
parameter describing the slope of the power spectrum is closer to what
is constrained by the data (n$_{\rm eff}$ rather than $\Omega_{\rm m}$
and n$_{\rm s}$).

We furthermore explored more physical priors for the evolution of the
temperature that do not allow sudden, large jumps in the temperature.
For reference, we have repeated the analysis in \cite{viel13WDM}) with
the reference priors (and thermal history parameterized as a power
law) used here and found that the the lower limit on $m_{\rm WDM}$
increases, from $>3.3\;\mathrm{keV}$ to $>3.9\;\mathrm{keV}$ (with the
same Planck priors on $\sigma_8$ and $n_{\rm s}$ as used by
\cite{viel13WDM}) and $>4.1\;\mathrm{keV}$ (with the reference priors
used in this work).

The Planck prior used puts a Gaussian prior of $\pm 0.01$ ($1\sigma$)
around Planck values for $\sigma_8$ and n$_{\rm eff}$.  The result
appears thus quite robust to changes in the choice of prior in
cosmological parameters and details of the analysis but is sensitive
to the assumed thermal priors.  If we drop the assumption of a
power-law evolution for the temperature of the IGM we get a lower
bound of $m_{\rm WDM} > 3.8\;\mathrm{keV}$.
  
This is a stronger limit than obtained by \citep{viel13WDM} in their
analysis of the HIRES/MIKE data, as well as in a more recent
reanalysis of the same data by \citep{garzilli15} that further relaxed
assumptions regarding the IGM temperature evolution. This is because
we have limited the change in temperature jumps between redshift bins
of $\Delta z=0.2$--$0.4$ to a physically plausible value of $\Delta
T=5000\rm\,K$.  Any change in the (volume averaged) IGM temperature
over $3<z<5.4$ will be due to either photo-heating of the IGM during
(QSO driven) HeII reionisation or adiabatic cooling due to the
expansion of the Universe.  Models that follow the expected HeII
ionising emissivity and spectral shape of quasars are consistent with
temperature boosts of at most $\Delta T=5000$--$10000\rm\,K$ which are
gradual and occur over $\Delta z \gtrsim 1$
\citep{puchwein15,upton,onorbe16}. The same holds for a plausible drop
in the temperature due to adiabatic cooling, which (ignoring heating)
will at most scale as $T_{0}\propto (1+z)^{2}$.  Larger values of
$\Delta T$ are not readily achievable within physically motivated
reionization models.  For this reason, \citep{viel13WDM} strongly
disfavored the binned analysis they performed for completeness that
weakened their constraint by around $1\rm\,keV$.

Our analysis of the combined data sets also gives significantly
strengthened constraints on the WDM mass, driven mostly again by the
high redshift HIRES/MIKE data set. However, unlike combining the above
data set with SDSS (as in \cite{viel13WDM} where the inclusion of
SDSS-II data did not impact on the free-streaming constraints), our
combined analysis of the high-redshift, high-resolution data with the
XQ-100 sample gives a significantly stronger lower limit on $m_{\rm
  WDM}$. This is again mostly due to the more physical temperature
evolution that we assumed when combining the two data sets, that does
not allow for sudden jumps in the temperature evolution. We expect
that were such a prior on the temperature evolution used also in the
case when combining SDSS-II with HIRES/MIKE, that a stronger bound on
the WDM mass would also be obtained.

Much like in our analysis of the HIRES/MIKE only data set, the results
for the combined data sets is -- apart from the thermal priors --
largely independent of the choices of prior and thus robust.  For the
combined data sets the $2\sigma$ C.L. lower limit is $5.3$ keV.  This
limit again weakens if we use the freely-floating temperature bins
rather than a power law to $3.5$ keV. The same priors were used in
this analysis as for the HIRES/MIKE only analysis, with an upper limit
on the difference in temperature between adjacent redshift bins again
of $\Delta T= 5000\rm\, K$.

\begin{table}[h]
\small
\begin{tabular}{lllll}%{|@{$\;\;$}r@{$\;$}|@{$\;$}c@{$\;$}|@{$\;$}c@{$\;$}|,@{$\;$}|}
\hline
Parameter & {\rm XQ-100} & {\rm HIRES/MIKE} & {\rm Combined}\\
\hline
$\mWDM\;\mathrm{[keV]}$ & $>1.4$& $>4.1$ & $>5.3$ \\
$\sigma_8$  & $[0.75,0.92]$& $[0.75,1.32]$ & $[0.83,0.95]$\\
$n_{\rm eff}$ & $[-2.42,-2.25]$& $[-2.53,-2.11]$& $[-2.43,-2.32]$\\ 
$T^A(z_{\rm p})\;\mathrm{[10^4\,K]}$& $[0.73,1.27]$& $[0.46,1.12]$ & $[0.74,1.06]$\\
$T^S(z_{\rm p})$ & $[-4.39,1.89]$& $[-4.78,-1.80]$ & $[-3.22,-0.82]$\\
$\gamma^A(z_{\rm p})$ & $[1.12,1.45]$& $[1.08,1.52]$& $[1.23,1.69]$\\
$\gamma^S(z_{\rm p})$ & $[-1.89,0.17]$& $[-1.18,1.77]$ & $[-0.07,1.81]$\\
$z_{\rm rei}$ & $[6.5,15.66]$& $[6.26,14.88]$ & $[6.25,13.43]$\\
$f_{\rm UV}$  & $[0.06,0.96]$& $[0.05,0.96]$ & $[0.05,0.94]$\\
$\chi^2/d.o.f.$ & $134/124$& $33/40$& $185/173$\\
\hline
\end{tabular}
\caption{Marginalized constraints at $95$ \%. Pivot redshifts for
  different data sets are: $z_p=3.6, 4.5, 4.2$ for XQ-100, HIRES/MIKE
  and combined.}
\label{tab:constraints}
\end{table}

\section{Conclusions}
\label{conclusec}

We have presented new constraints on the free-streaming of WDM based
on an MCMC analysis of the XQ-100 and HIRES/MIKE \lya\ forest data
sets. The new constraints in terms of the mass of a thermal relic WDM
particle, $m_{\rm WDM} > 5.3\;\mathrm{keV}$ at 2$\sigma$, are the
strongest to date, and thus imply significantly colder dark matter
than the $2-2.5$ keV values that are typically required to mitigate
tensions in the predictions of cold dark matter models with other
astrophysical observations on small scales.

Previous analysis of the same high-resolution \lya\ forest data had
given constraints of $m_{\rm WDM} > 3.3\;\mathrm{keV}$
(\citep{viel13WDM} using HIRES/MIKE) and $>3.95\;\mathrm{keV}$
(\citep{baur15}; using SDSS-III/BOSS). Adding the new data from the
XQ-100 survey which has similar redshift coverage as SDSS, but extends
to significantly smaller scales, has strengthened the constraints to a
significantly smaller free-streaming length and corresponding larger
values of the mass of a thermal relic WDM particle.  Another important
aspect of our new analysis was the assumption of more physical priors
on the gas temperature evolution with redshift.  While the results of
our analysis for the new XQ-100 data alone give relatively weak
constraints, the combined analysis is very robust to different choices
of priors and also gives a largely consistent picture with
independent, more direct measurements of the the thermal history of
the IGM over a wide redshift range $z=3 - 5.4$ \citep{becker11}.  This
is due to the complementarity of the XQ-100 and MIKE/HIRES data sets
in redshift and the corresponding longer ``lever arm'' of the combined
sample in redshift.  On large scales the main degeneracies are between
the mean flux, gas temperature and the amplitude of matter
fluctuations.  At the lower redshifts probed by the moderate
resolution XQ-100 data, the WDM constraints are coming predominantly
from these scales and are therefore coupled to the same degeneracies.
On small scales, however, the cut-off in the flux power spectrum is
dominated by thermal, Jeans or WDM smoothing, which introduces a
different set of degeneracies.  At higher redshifts the WDM smoothing
scale increases relative to the thermal/Jeans smoothing scale in
velocity space, thus most constraining power comes from small scales
at high redshift.

%A simple picture of the how different scales dominate the cutoff in
%the power spectrum suggests that while it is always beneficial to
%measure the power at as small scales as possible, lower redshift
%measurements benefit from smaller scales the most; while high redshift
%measurements in their current state would benefit mostly from more
%QSOs. Degeneracies between the parameters play a twofol role: on large
%scale the main degeneracy is between amplitude of the mean flux and
%temperature and amplitude of matter fluctuations. At lower redshifts
%WDM constraints are coming predominantly from these scales and are
%coupled to the same degeneracies. On small scales the cutoff in flux
%power is dominated by thermal, Jeans (or WDM) smoothing, which
%introduces degeneracies between the parameters controlling those
%cutoff scales. At higher redshifts WDM smoothing scale becomes
%comparable to the other two scales, thus most constraining power is
%coming from small scales.  

We conclude with three important remarks. First, the tightest limit
presented here weakens if we drop the assumption of a power-law
evolution for the temperature and use instead a thermal history with
freely floating $T(z)$ bins (but with limited temperature jumps
between adjacent bins) in our analysis.  In this case the limit
weakens to $3.5$ keV, a number which is very similar to the one
obtained from similar analyses of HIRES/MIKE data assuming a power-law
evolution of the thermal parameters characterising the IGM
\citep{viel13WDM}. Models with free-streaming lengths larger than this
require, however, an unphysical heating and/or cooling of the IGM over
a very short short timescale, in strong disagreement with theoretical
expectations for the IGM thermal history and measurements of the IGM
temperature based on high-resolution \lya forest data.

Secondly, \citep{hui16} have recently suggested that temperature
fluctuations could compensate for the WDM cutoff by providing an
increase of power at small scales (but see also \citep{lai06}).  This
is potentially an important systematic effect that should be better
quantified by performing template fitting based on more accurate
modelling of spatial fluctuations of the meta-galactic UV background,
as well as the residual temperature fluctuations from hydrogen
reionization with radiation hydrodynamical simulations that
incorporate radiative transfer effects rather than analytical
modeling.

Thirdly, it should (at least in principle) be possible to further
(moderately) strengthen the limits on the free-streaming of warm dark
matter by reducing the statistical errors of the high-redshift, small
scale flux power spectrum obtainable with high-resolution
spectrographs and further constraining the thermal and reionization
history of the IGM.

\appendix
\renewcommand\thefigure{\thesection.\arabic{figure}}
\setcounter{figure}{0}
\section{Parameter degeneracies}

\begin{figure*}
\begin{center}
\includegraphics[width=1.0\textwidth]{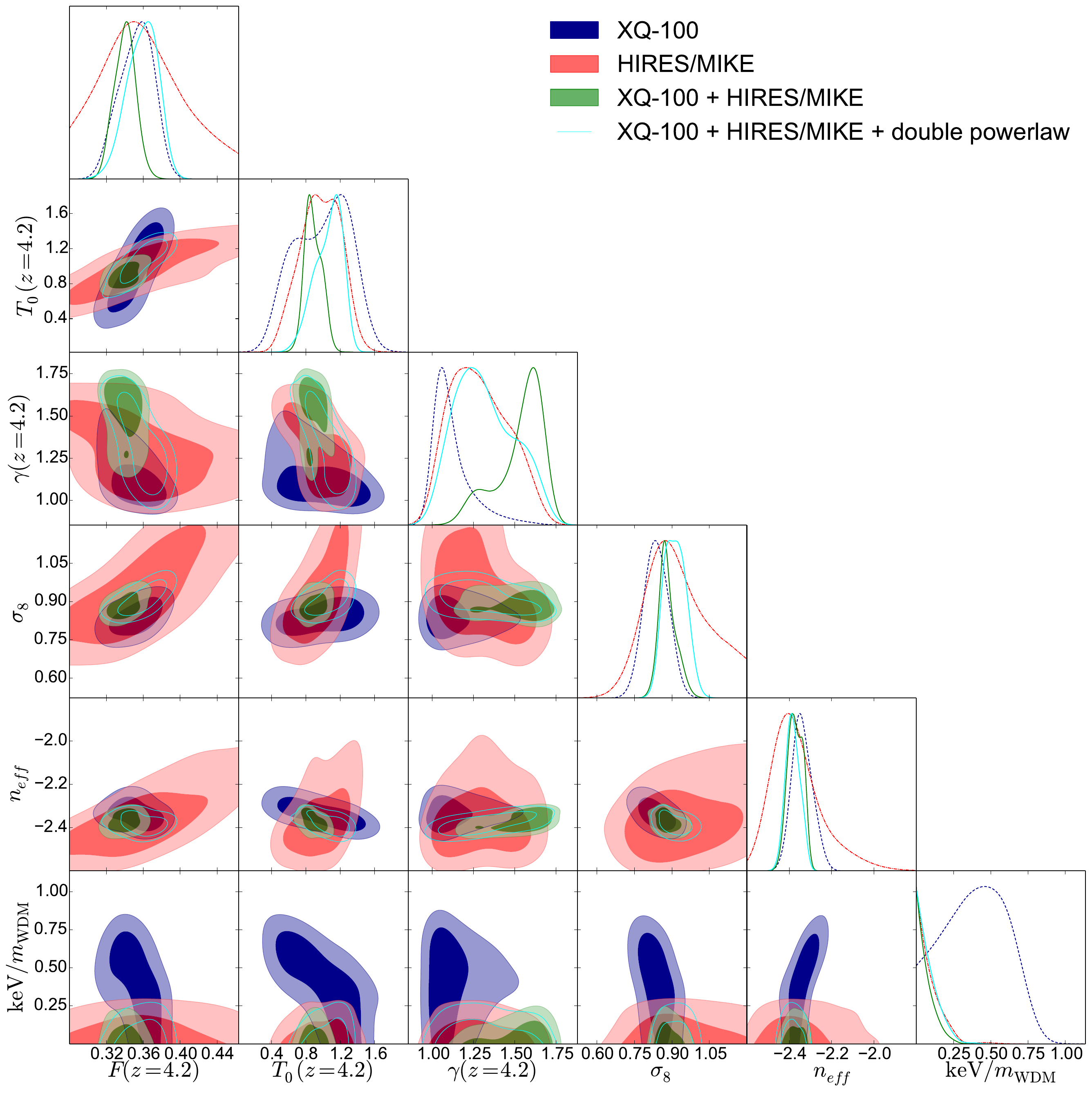}
\caption{Two dimensional posterior (marginalized) likelihood
  distributions for the main parameters for the XQ-100, HIRES/MIKE and
  combined data sets (blue, red and green curves), respectively. We
  also show contours when using a double power-law evolution of the
  thermal parameters (cyan curves), as described in more details in
  the text. Instead of $4$ parameters describing the thermal history
  only values evaluated at a specific redshift were chosen. The
  redshift chosen is where different data sets
  overlap.\label{fig_WDM}}
\end{center} 

\end{figure*}

Degeneracies between the parameters play an important role in how well
a specific parameter (e.g. free streaming length/WDM mass) can be
estimated using different data sets.  In the bottom row of
Fig.~\ref{fig_WDM}, it is clear that for low redshift data (XQ-100;
blue coloured contours), there are strong degeneracies between the
mass of a thermal relic WDM particle and the temperature (at a given
redshift).  This is not surprising, since both the temperature and WDM
effects change the power spectrum on large as well as small scales. At
lower redshifts the effects are small in both cases and thus harder to
distinguish within the observational errorbars. The degeneracy with
temperature is an anti-correlation that is expected; the data prefers
either higher temperatures and lower masses of WDM, or lower
temperatures with higher WDM masses.

However, whereas the temperature degeneracy with the free streaming
length comes as no surprise, the mean flux degeneracy might not be
naively expected (bottom left panel of Fig.~\ref{fig_WDM}).  Since
this degeneracy has a similar anti-correlation with the mass of the
WDM as seen for the temperature (also shown as positive correlation
between mean flux and temperature - top left 2D panel in
Fig.~\ref{fig_WDM}), it means the sensitivity of the XQ-100 data to
the WDM mass comes mostly from the overall amplitude of the flux power
spectrum, rather than its shape in the cutoff regime at smaller
scales.  A possible solution (apart from measuring different
statistics, and increasing the precision of the current measurements)
would be to increase the maximum scale up to which the flux power
spectrum is measured.  If the thermal/Jeans smoothing and smoothing
due to a high WDM mass are different enough a feature (kink) should be
observable on some (arbitrarily) small scales where the flux power
spectrum cutoff transitions from being dominated by the thermal/Jeans
smoothing to being dominated by the mass of the WDM. This would,
however, only work if the WDM mass is large enough.
%(at lower redshifts).

The above degeneracies almost disappear when using the higher redshift
data in the analysis (HIRES/MIKE; red coloured contours).
Fig.~\ref{fig_WDM} shows no appreciable degeneracy between mass of the
WDM and any other parameters. This is because at higher redshifts, for
the WDM masses we consider here the cutoff scale by the
free-streaming of the WDM becomes more and more important and this
scale will show no redshift evolution and will be thus easier to pick
up in the data. This is the reason why the higher
redshift data becomes such a powerful tool for constraining the
free-streaming. To increase the constraining power, more observations
to decrease the statistical errors would be more beneficial than
pushing to smaller scales (although the latter would be helpful as
well).  This is because the MCMC analysis shows that the constraints
on the free-streaming length are largely independent of the different
assumed values of priors, meaning that the resulting lower bound on
the mass of the WDM is driven by the statistical error.

Lastly we draw attention to a slight discrepancy in the measurement of
the slope of the $T-\rho$ relation (third row of
Fig.~\ref{fig_WDM}). The value of $\gamma(z=4.2)$ measured from the
low and high redshift data sets are in modest ($1-2\sigma$) tension.
While this could be a statistical fluke, we would like to point out
that this might actually be an expected result, if HeII reionization
happens somewhere between redshift $3-4$. The high redshift data
(HIRES/MIKE) measures the thermal history above a redshift of $z=4.2$,
where HeII reionization (that happens at lower redshifts) would have
little effect.  The value of $\gamma$ at these redshifts is thus
expected to slowly increase and approach the asymptotic value of
around $1.6$ \citep{schaye00,puchwein15,upton}).  The evolution is
well described by a power-law in redshift.  However, with HeII
reionization somewhere between $z=3 - 4$, a feature is to be expected
in the evolution of $\gamma$, where its value falls to
$\gamma=1.2$--$1.3$ and then start to rise again towards higher
asymptotic values over a redshift interval of $\Delta z \sim 2$
\cite{puchwein15}.  If we fit such a feature with a simple power-law
in redshift, a lower amplitude for the power-law would be obtained
compared to the case where there is no feature in the evolution, and
no HeII reionization.  This is what the data is preferring - higher
values of $\gamma$ measured from high redshift data set, and lower
overall amplitude of $\gamma$ at lower redshifts.  We note, however,
that a more detailed model of $\gamma(z)$ evolution may be necessary
at lower redshift to capture possible HeII reionization effects.
Dropping the assumption of a simple power-law describing the evolution
of $T_0(z)$ and $\gamma(z)$, and allowing for the power-law evolution
to have different slope below and above $z_{p} = 4.2$, relaxes this
tension considerably, as is shown in the third row of
Fig.~\ref{fig_WDM} (magenta colour -- double powerlaw). The WDM
constraints in this case are slightly weaker compared to the reference
case of the analysis of the combined data sets, and exclude WDM masses
above $m_{\rm WDM} > 4.5$ keV.  The tension disappears for the case of
using $T_0$ in independent redshift bins, even though $\gamma(z)$ is
still described as a single power-law in such a case. The WDM limits
derived from this case are described in the main text.

\section{Degeneracy between WDM mass and redshift of reionization}
\label{app:B}

Due to the fact that smoothing from both WDM thermal relic as well as
pressure smoothing act on the 3D matter power spectrum a certain
amount of degeneracy between the parameters is expected. However, in
this section we show that this degeneracy is largely broken by the
long redshift range considered in the data analysis.

\begin{figure}
\begin{center}
\includegraphics[width=1.0\linewidth]{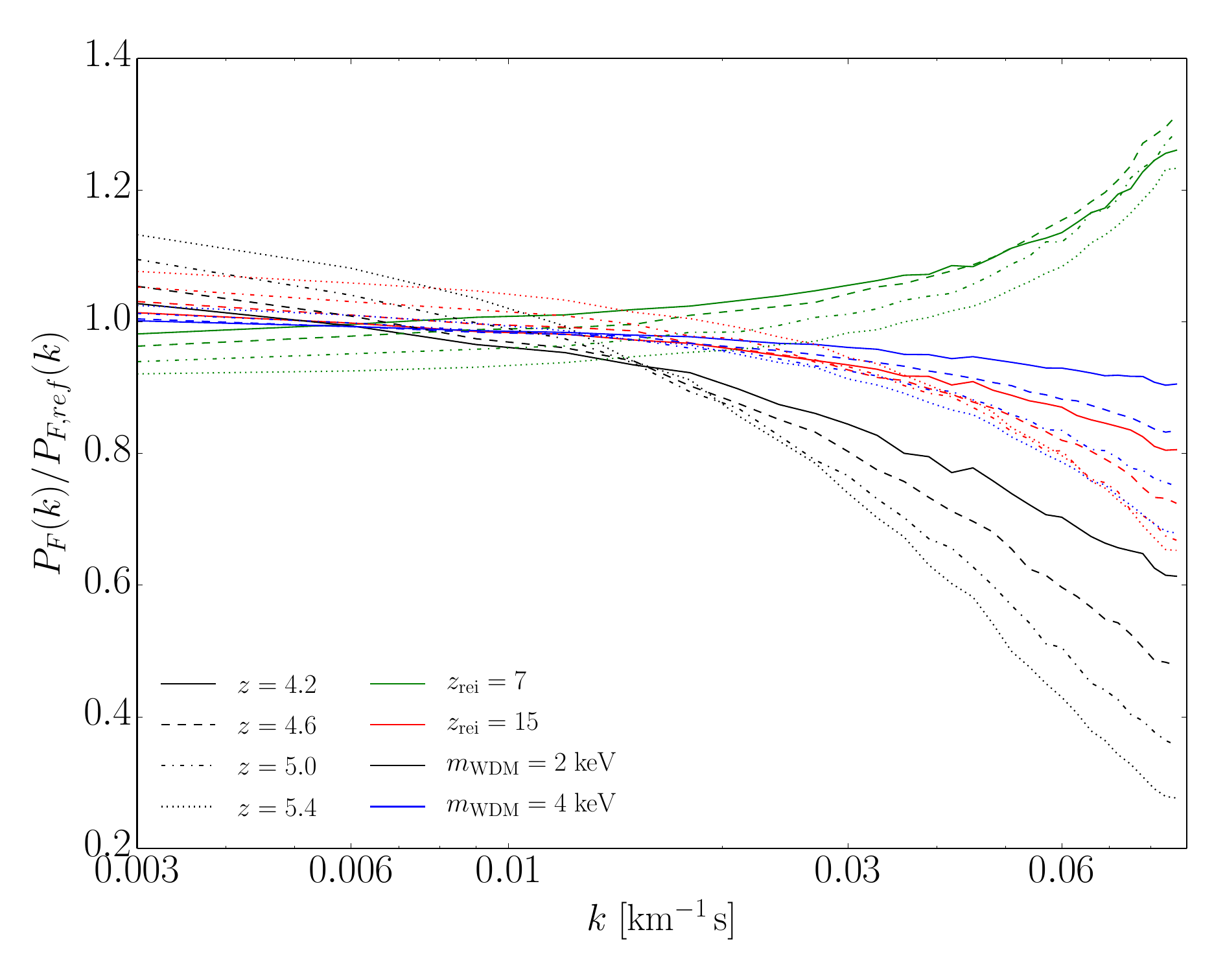}
\caption{The flux power spectrum for different models varying mass of
  the WDM ($m_{\rm WDM}$) and redshift of reionization ($z_{\rm
    rei}$). The colours show two values of $m_{\rm WDM}$ -
  2$\;\mathrm{keV}$ in blue and 4$\;\mathrm{keV}$ in black - and two
  values of $z_{\rm rei}$ - 7 in green and 15 in red. The reference
  model against which the flux power is compared, was $\Lambda$CDM
  model with $z_{\rm rei} = 9$. Different line styles show the
  redshift evolution of the flux power: full line ($z=4.0$), dashed
  line ($z=4.6$), dot-dashed line ($z=5.0$) and dotted line
  ($z=5.4$). Only the effect on the largest redshift bins is shown
  since that is where the effect of WDM on the smoothing scale
  becomes largest. The plot illustrates how the redshift evolution
  changes the shape of the flux power differently for the two parameters.\label{fig_pkzr}}
\end{center} 
\end{figure}

Fig.~\ref{fig_pkzr} shows the flux power ratio when we vary WDM
and $z_{\rm rei}$ models compared to the reference $\Lambda$CDM
case. 
The plot nice nicely illustrates how the different redshift evolution
of the effect of reionization redshift and free-streaming of the dark
matter on the flux power spectrum makes it possible to separate the
two effects. To
fully capture the effect these two degenerate parameters have, we used
used a grid of simulations that samples the parameter plane of
$1/m_{\rm WDM}$ and $z_{\rm rei}$.

\begin{figure}
\begin{center}
\includegraphics[width=1.0\linewidth]{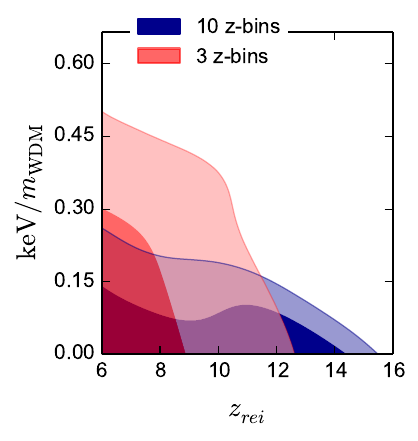}
\caption{The 2D posterior likelihood contours in the parameter plane
  of mass of WDM particle and redshift of reionization. Different colours
  represent different subsets of the combined data set used. In
  particular, the blue colour shows the full analysis of the combined
  data (XQ-100 + HIRES/MIKE) which used 10 redshift bins. In red
  we show the results when only 3 redshift bins were used in the
  analysis, centered around $z=4.2$ ($z=4.0,4.2,4.6$).\label{fig_2dzr}}
\end{center} 
\end{figure}

To illustrate the effect of redshift evolution further, we show a 2D
plot of the posterior likelihood distribution in the parameter plane
of $m_{\rm WDM}$ and $z_{\rm rei}$ (Fig.~\ref{fig_2dzr}). The
degeneracy between the two parameters is increased when only three
redshift bins are considered in the analysis. These redshift bins also
do not span the whole length of the redshift range the combined data
set tracks, but are centered around the pivot redshift of $z=4.2$
($z=4.0,4.2,4.6$).

\begin{figure}
\begin{center}
\includegraphics[width=1.0\linewidth]{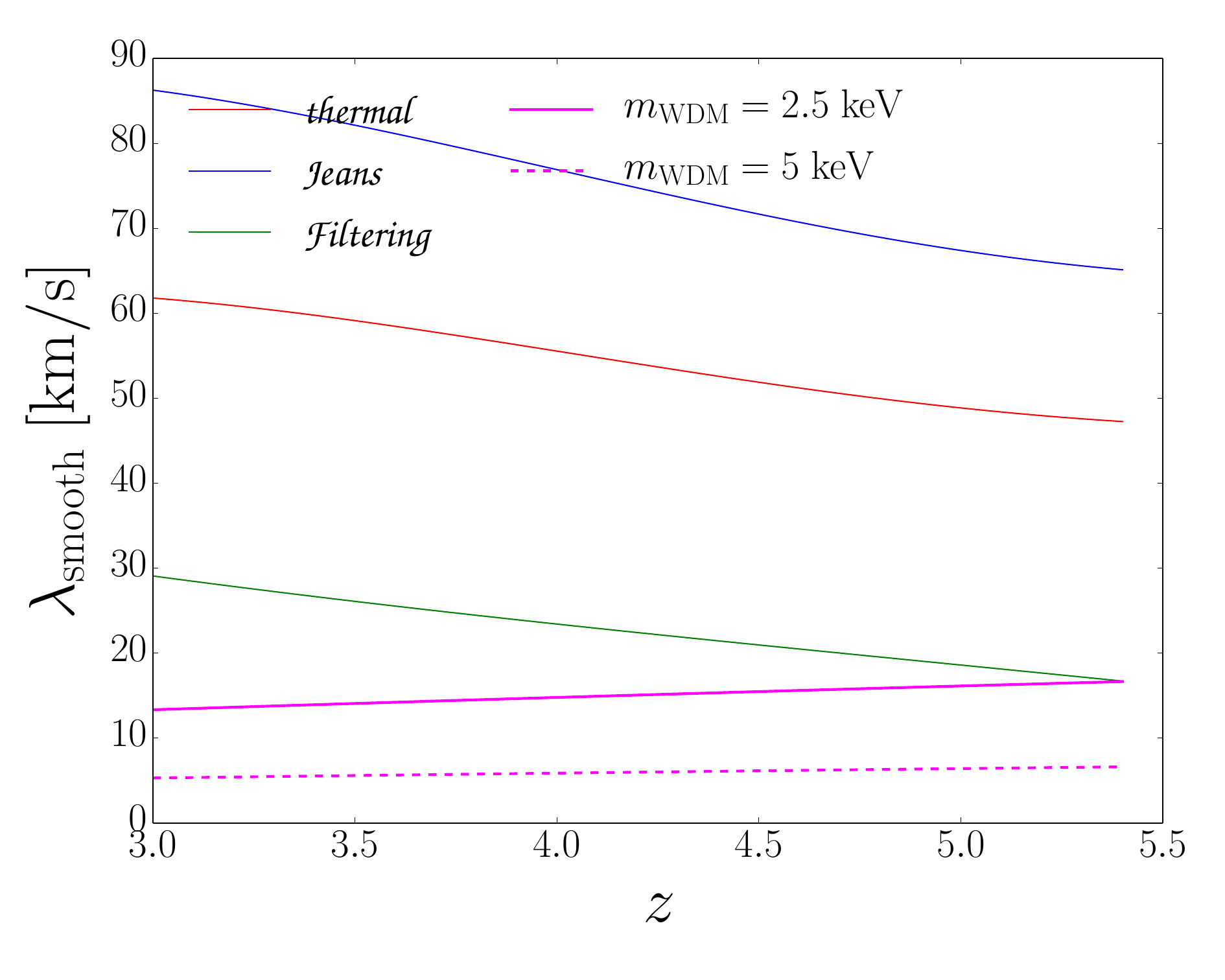}
\caption{The redshift evolution of the different smoothing scales in
  units of $\kms$: thermal
  (red), Jeans (blue), filtering (green) and free-streaming from WDM
  thermal relic (magenta). The two line-styles show different values of the
  WDM mass for 2.5$\;\mathrm{keV}$ (full line) and 5$\;\mathrm{keV}$
  (dashed line) respectively. 
  %% The arrows on the right-hand side of
  %% the plot indicate the maximum scale ($\sim 2\pi/k_{\rm max}$) for
  %% three measurements of the Lyman-$\alpha$ forest flux power spectrum.
  \label{fig_jeans}}
\end{center} 
\end{figure}

Furthermore, in Fig.~\ref{fig_jeans} we show the redshift evolution of
the different smoothing scales that affect the cut-off in the flux
power spectrum. We have used the thermal history evolution of our
reference model in the case of thermal, Jeans and filtering
scale. The filtering scale shown on this plot was computed using the relation
in \citep{gnedin98} - i.e. $\lambda_F = 2\pi/k_F$. 
We also show the Jeans smoothing scale since it has
been argued in \citep{gnedin98} that the filtering scale $\lambda_F$
will always be smaller than the Jeans scale $\lambda_J$. Thus the
Jeans scale plays a role of a (conservative) upper limit on the amount
of pressure smoothing. The plot shows that thermal and
filtering (or Jeans) scales have a very different redshift evolution
compared to the free-streaming scale of the warm dark matter, which is
the only scale slowly increasing with redshift in velocity
space. Fig.~\ref{fig_jeans} is meant to be of illustrative purpose
only, to show that different scales evolve differently with
redshift. We would also like to caution the reader that, while the
pressure smoothing scale (Jeans or filtering) and WDM free-streaming
scale, are acting on the 3D matter density field, the thermal scale is
a 1D smoothing scale that operates on the optical depth field.
%% which remains constant in comoving coordinates.

\begin{figure}
\begin{center}
\includegraphics[width=1.0\linewidth]{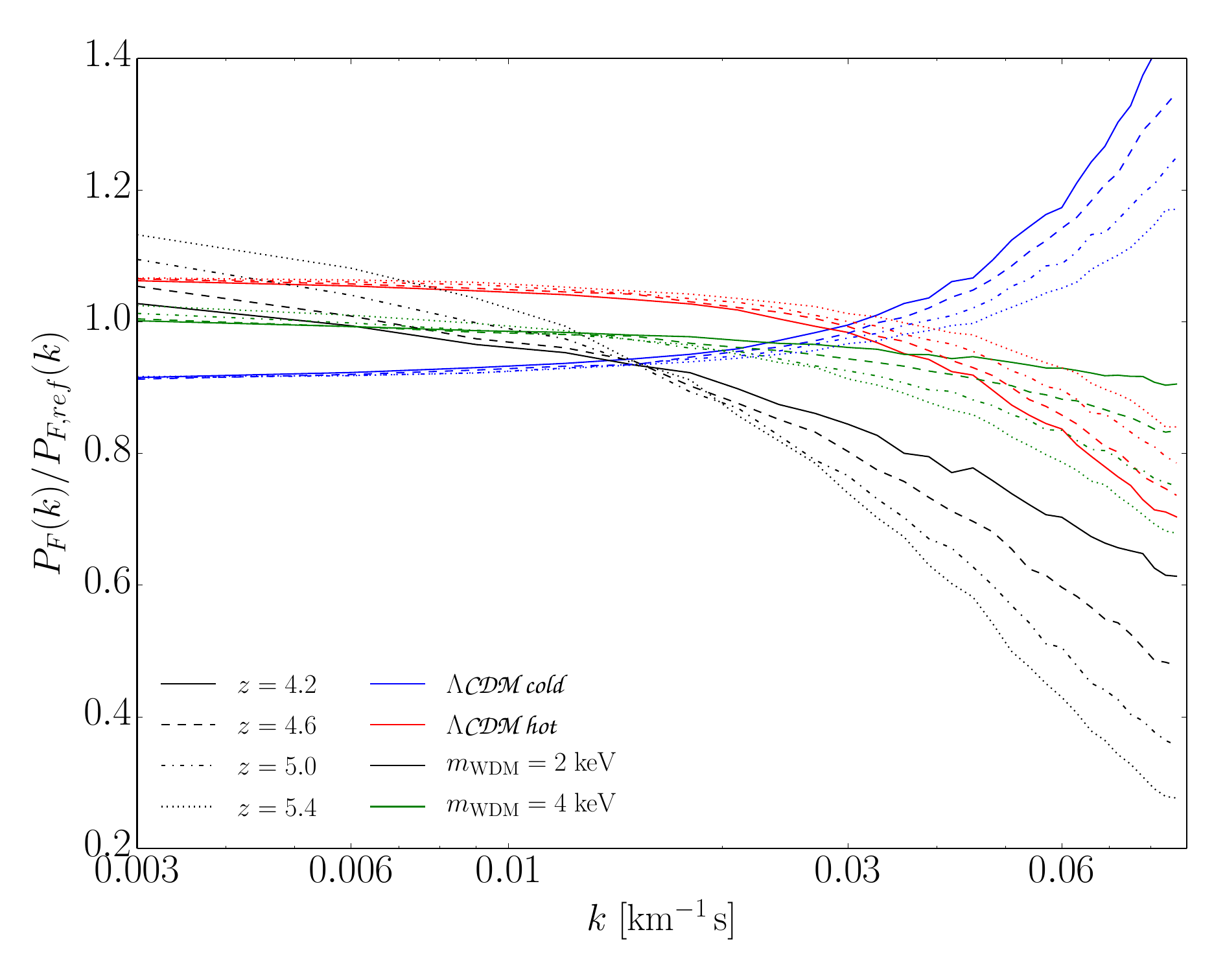}
\caption{The flux power spectrum for different models varying mass of
  the WDM ($m_{\rm WDM}$) and the amplitude of the IGM temperature at
  the mean density ($T_0^A$). The colours show two values of $m_{\rm
    WDM}$ - 2$\;\mathrm{keV}$ in black and 4$\;\mathrm{keV}$ in green - and two
  $\Lambda$CDM models with different temperatures - hot IGM in red (roughly
  $3000$K hotter) and cold in blue (roughly $3000$K colder). The reference
  model against which the flux power is compared, was $\Lambda$CDM
  model with $T_0(z=3.6) = 11000$K. Different line styles show the
  redshift evolution of the flux power: full line ($z=4.0$), dashed
  line ($z=4.6$), dot-dashed line ($z=5.0$) and dotted line
  ($z=5.4$). As with Fig.~\ref{fig_pkzr}, only the effect on the
  largest redshift bins is shown since that is where the effect of WDM
  on the smoothing scale becomes largest.\label{fig_pktemp}}
\end{center} 
\end{figure}

While Fig.~\ref{fig_jeans} shows that the redshift evolution differs
between different smoothing scales, the MCMC bounds derived in this
paper make use of the full shape of the flux power spectrum. Furthermore,
the flux power spectrum traces the integral over the 3D matter power,
and is thus sensitive to small scales at any given parallel
wave-number. Indeed, this is why lower resolution surveys are also
able to put bounds on the WDM free-streaming scale
\citep{baur15}. The effect of redshift evolution on the shape of the
flux power spectrum is shown in Fig.~\ref{fig_pktemp}. Even on large
scales ($k < 0.01\skm$), models with varying amount of thermal or WDM
free-streaming smoothing have quite distinct shapes. Combining the
shape with the redshift evolution helps break the degeneracies among
the IGM parameters and the mass of the WDM.

\section{The effect of different priors on the limits of $m_{\rm
    WDM}$}

In this section we show an extended table of how the limits on the
mass of the WDM change when imposing different priors. The priors
for reference case and weak priors are the same as the ones plotted in
Fig.~\ref{fig_3}. 

The reference priors consist of weak priors on
the values of mean flux in each redshift bin. These priors were chosen
to be Gaussian with mean value as predicted by the empirical fit by
\citep{palanque13} with $\pm$ 0.04 (1$\sigma$). Further, the reference
priors include bounds on some of the parameters that are physically
motivated: $m_{\rm WDM} \ge 0$, $6 \leq z_{\rm rei} \leq 16$,
$0\;\mathrm{K} \leq T_0^A \leq 20000\;\mathrm{K}$, $-5 \leq T_0^S \leq
5$ and $1 \leq \gamma(z_i) < 1.7$ for each redshift $z_i$. We have
also checked that the exact values for upper and lower bounds on
$z_{\rm rei}$ and the temperature amplitude ($T_0^A$) and slope
($T_0^S$) do not have an impact on the final constraints of the
WDM. The bounds for $\gamma$ at each redshift are physically motivated
for the time of HeII reionization \citep{mcquinn09,mcquinn15}.

\begin{table}[h]
\small
\begin{tabular}{l|llll}%{|@{$\;\;$}r@{$\;$}|@{$\;$}c@{$\;$}|@{$\;$}c@{$\;$}|,@{$\;$}|}
\hline
Priors used & {\rm XQ-100} & {\rm HIRES/MIKE} & {\rm Combined}\\
\hline
Reference & $>1.38$& $>4.08$ & $>5.32$ \\
Weak priors & $>1.33$& $>4.11$ & $>5.21$ \\
Planck priors & $>1.34$& $>3.95$ & $>5.25$ \\
Thermal priors & $>1.49$& $>2.11$ & $>3.48$ \\
\hline
\end{tabular}
\caption{Marginalized constraints at $95$ \% for $m_{\rm WDM}$ in the
  units of keV. Different priors used are the reference case, weak
  priors, Planck priors on the cosmological parameters and physical
  priors on thermal evolution where $T_0$ varies freely with redshift
  bins. Compared to the result shown in Table~\ref{tab:constraints}
  more decimal points are shown in the result.}
\label{tab:priors}
\end{table}

The weak priors, as already described in the main body of the text,
add the following priors to the reference values: $\sigma_8$ and n$_{\rm
  eff}$ have Gaussian priors of $\pm$ 0.1 (1$\sigma$) around Planck
values and the assumed temperature $T_0^A$ is $10000 \pm 5000$ K
(1$\sigma$). 

Furthermore, priors on cosmological parameters were added to the
reference once (Planck priors), such that: $\sigma_8$ and n$_{\rm
  eff}$ have Gaussian priors of $\pm$ 0.01 (1$\sigma$) around Planck
values.

And lastly, we also considered a temperature evolution where the
temperature $T_0$ was allowed to vary freely in each of the redshift
bins. In this case we have used references values for priors to which
we have added additional constraint on the change of the temperature
between redshift bins, such that the change in temperature jumps
between redshift bins of $\Delta z=0.2$--$0.4$ was limited to a
physically plausible value of $\Delta T=5000\rm\,K$.

\section{Bestfit and confidence levels tables}

In this section we show the full tables of the best-fit parameters (and
their $1$ and $2\sigma$ confidence intervals) for the MCMC analysis of
the three different data sets: XQ-100 (Table~ S1), HIRES/MIKE (Table~
S2) and combined XQ-100 + HIRES/MIKE (Table~S3).

\begin{table}[h]
\begin{tabular}{llllll}%{lllllllll}%{|@{$\;$}r@{$\;$}|@{$\;$}c@{$\;$}|@{$\;$}c@{$\;$}|@{$\;$}|}
\hline
Parameter & $(1\sigma)$ & $(2\sigma)$ & Best fit \\
\hline
${\bar F}(z=3.0)$ & $[0.67,0.68]$ & $[0.65,0.70]$ & $0.68$ \\
${\bar F}(z=3.2)$ & $[0.61,0.62]$ & $[0.59,0.64]$ & $0.61$ \\
${\bar F}(z=3.4)$ & $[0.54,0.56]$ & $[0.53,0.57]$ & $0.55$ \\
${\bar F}(z=3.6)$ & $[0.50,0.51]$ & $[0.49,0.53]$ & $0.51$ \\
${\bar F}(z=3.8)$ & $[0.44,0.45]$ & $[0.41,0.46]$ & $0.44$ \\
${\bar F}(z=4.0)$ & $[0.38,0.39]$ & $[0.36,0.41]$ & $0.39$ \\
${\bar F}(z=4.2)$ & $[0.34,0.36]$ & $[0.32,0.38]$ & $0.35$ \\
$T_0^A(z=z_p)\;\mathrm{[10^4\,K]}$ & $[0.97,1.12]$ & $[0.73,1.27]$ & $1.03$ \\
$T_0^S(z=z_p)$ & $[-1.54,0.73]$ & $[-4.39,1.89]$ & $-0.69$
\\
$\gamma^A(z=z_p)$ & $[1.25,1.31]$ & $[1.12,1.45]$ & $1.28$ \\
$\gamma^S(z=z_p)$ & $[-1.55,-0.97]$ & $[-1.89,0.17]$ &
$-1.14$ \\
$\sigma_8$ & $[0.81,0.86]$ & $[0.75,0.92]$ & $0.84$ \\
$z_{\rm rei}$ & $[9.92,13.47]$ & $[6.50,15.66]$ & $11.50$ \\
$n_{\rm eff}$ & $[-2.36,-2.32]$ & $[-2.42,-2.25]$ & $-2.34$ \\
$1/m_{\rm WDM}\;[\mathrm{keV^{-1}}]$ & $[0,0.63]$ & $[0,0.77]$ &
$0.40$ \\
$f_{\rm UV}$ & $[0.36,0.72]$ & $[0.06,0.96]$ & $0.53$ \\
\hline
\end{tabular}
\caption{XQ-100: Marginalized constraints at $68$ and $95$ \%,
  obtained from the MCMC analysis as well as bestfit values. The pivot
  redshift is $z_{\rm p}=3.6$.}
\label{tab:xq}
\end{table}

\begin{table}[h]
\begin{tabular}{lllll}%{|@{$\;\;$}r@{$\;$}|@{$\;$}c@{$\;$}|@{$\;$}c@{$\;$}|,@{$\;$}|}
\hline
Parameter & $(1\sigma)$ & $(2\sigma)$ & Best fit \\
\hline
${\bar F}(z=4.2)$ & $[0.33,0.38]$ & $[0.27,0.46]$ & $0.36$ \\
${\bar F}(z=4.6)$ & $[0.25,0.29]$ & $[0.21,0.37]$ & $0.27$ \\
${\bar F}(z=5.0)$ & $[0.14,0.17]$ & $[0.12,0.21]$ & $0.16$ \\
${\bar F}(z=5.4)$ & $[0.04,0.06]$ & $[0.03,0.08]$ & $0.05$ \\
$T_0^A(z=z_p)\;\mathrm{[10^4\,K]}$ & $[0.71,0.91]$ & $[0.46,1.12]$ &
$0.80$ \\
$T_0^S(z=z_p)$ & $[-3.93,-3.11]$ & $[-4.78,-1.80]$ &
$-3.46$ \\
$\gamma^A(z=z_p)$ & $[1.22,1.38]$ & $[1.08,1.52]$ & $1.30$ \\
$\gamma^S(z=z_p)$ & $[-0.28,0.81]$ & $[-1.18,1.77]$ & $0.25$
\\
$\sigma_8$ & $[0.86,1.01]$ & $[0.75,1.32]$ & $0.96$ \\
$z_{\rm rei}$ & $[8.44,11.52]$ & $[6.26,14.88]$ & $10.12$ \\
$n_{\rm eff}$ & $[-2.43,-2.33]$ & $[-2.53,-2.11]$ & $-2.36$ \\
$1/m_{\rm WDM}\;[\mathrm{keV^{-1}}]$ & $[0,0.17]$ & $[0,0.28]$ &
$0.09$ \\
$f_{\rm UV}$ & $[0.35,0.70]$ & $[0.05,0.96]$ & $0.52$ \\
\hline
\end{tabular}
\caption{HIRES/MIKE: Marginalized constraints at $68$ and $95$ \%,
  obtained from the MCMC analysis as well as bestfit values. The pivot
  redshift is $z_{\rm p} = 4.5$.}
\label{tab:mh}
\end{table}

\newpage

\begin{table}[h]
\begin{tabular}{lllll}%{|@{$\;\;$}r@{$\;$}|@{$\;$}c@{$\;$}|@{$\;$}c@{$\;$}|,@{$\;$}|}
\hline
Parameter & $(1\sigma)$ & $(2\sigma)$ & Best fit \\
\hline
${\bar F}(z=3.0)$ & $[0.69,0.70]$ & $[0.67,0.71]$ & $0.69$ \\
${\bar F}(z=3.2)$ & $[0.62,0.63]$ & $[0.61,0.64]$ & $0.63$ \\
${\bar F}(z=3.4)$ & $[0.56,0.57]$ & $[0.54,0.58]$ & $0.56$ \\
${\bar F}(z=3.6)$ & $[0.51,0.52]$ & $[0.50,0.53]$ & $0.52$ \\
${\bar F}(z=3.8)$ & $[0.45,0.46]$ & $[0.44,0.47]$ & $0.45$ \\
${\bar F}(z=4.0)$ & $[0.38,0.39]$ & $[0.37,0.40]$ & $0.38$ \\
${\bar F}(z=4.2)$ & $[0.33,0.35]$ & $[0.32,0.36]$ & $0.34$ \\
${\bar F}(z=4.6)$ & $[0.25,0.27]$ & $[0.23,0.29]$ & $0.26$ \\
${\bar F}(z=5.0)$ & $[0.13,0.14]$ & $[0.11,0.17]$ & $0.13$ \\
${\bar F}(z=5.4)$ & $[0.03,0.04]$ & $[0.01,0.06]$ & $0.04$ \\
$T_0^A(z=z_p)\;\mathrm{[10^4\,K]}$ & $[0.83,0.94]$ & $[0.74,1.06]$ &
$0.89$ \\
$T_0^S(z=z_p)$ & $[-2.59,-1.99]$ & $[-3.22,-0.82]$ & $-2.23$
\\
$\gamma^A(z=z_p)$ & $[1.51,1.63]$ & $[1.23,1.69]$ & $1.53$ \\
$\gamma^S(z=z_p)$ & $[0.81,1.42]$ & $[-0.07,1.81]$ & $1.04$
\\
$\sigma_8$ & $[0.87,0.89]$ & $[0.83,0.95]$ & $0.88$ \\
$z_{\rm rei}$ & $[8.21,11.37]$ & $[6.25,13.43]$ & $9.80$ \\
$n_{\rm eff}$ & $[-2.39,-2.35]$ & $[-2.43,-2.32]$ & $-2.37$ \\
$1/m_{\rm WDM}\;[\mathrm{keV^{-1}}]$ & $[0,0.13]$ & $[0,0.22]$ &
$0.07$ \\
$f_{\rm UV}$ & $[0.32,0.65]$ & $[0.05,0.94]$ & $0.48$ \\
\hline
\end{tabular}
\caption{Combined: Marginalized constraints at $68$ and $95$ \%,
  obtained from the MCMC analysis as well as bestfit values. The pivot
  redshift is $z_{\rm p} = 4.2$.}
\label{tab:al}
\end{table}

%\vskip 10pt

\bigskip
\acknowledgments VI is supported by US NSF grant AST-1514734. VI also
thanks M. McQuinn for useful discussions, and IAS, Princeton, for
hospitality during his stay where part of this work was completed. MV
and TSK are supported by ERC-StG "cosmoIGM". SL has been supported by
FONDECYT grant number 1140838 and partially by PFB-06 CATA. VD, MV, SC
acknowledge support from the PRIN INAF 2012 "The X-Shooter sample of
100 quasar spectra at $z \sim 3.5$: Digging into cosmology and galaxy
evolution with quasar absorption lines. GB is supported by the NSF
under award AST-1615814. SLE acknowledges the receipt
of an NSERC Discovery Grant. MH acknowledges support by ERC ADVANCED
GRANT 320596 "The Emergence of Structure during the epoch of
Reionization". LC is supported by YDUN DFF – 4090-00079. KDD is supported by an NSF AAPF
fellowship awarded under NSF grant AST-1302093. JSB acknowledges the
support of a Royal Society University Research Fellowship. Based on observations
collected at the European Organisation for Astronomical Research in
the Southern Hemisphere under ESO programme 189.A-0424. This work
made use of the DiRAC High Performance Computing System (HPCS) and the
COSMOS shared memory service at the University of Cambridge. These are
operated on behalf of the STFC DiRAC HPC facility. This equipment is
funded by BIS National E-infrastructure capital grant ST/J005673/1 and
STFC grants ST/H008586/1, ST/K00333X/1.
%\end{acknowledgments}

\bibliographystyle{unsrt}
\bibliography{Bibliofile}

%\pagebreak

%{\center{\Large{\textsc{Supplemental Material}}}}

%\section{Results}

\end{document}